\documentclass[aps,twocolumn,showpacs]{revtex4}
\usepackage{graphics}

\begin{document}

\title{Bulk gravitational field and dark radiation on the brane 
in dilatonic brane world}

\author{Hiroyuki Yoshiguchi}
\address{Department of Physics, University of Tokyo 7-3-1 Hongo,
Bunkyo, Tokyo 113-0033, Japan}

\author{Kazuya Koyama}
\address{Department of Physics, University of Tokyo 7-3-1 Hongo,
Bunkyo, Tokyo 113-0033, Japan}

\begin{abstract}
We discuss the connection between the dark radiation on the brane 
and the bulk gravitational field in a dilatonic brane world model proposed
by Koyama and Takahashi where the exact solutions for the five
dimensional cosmological perturbations can be obtained 
analytically. It is shown that the dark radiation perturbation 
is related to the non-normalizable Kaluza-Klein (KK) mode of 
the bulk perturbations.  
For the de Sitter brane in the anti-de Sitter bulk, the 
squared mass of this KK mode is $2 H^2$ where $H$ is the 
Hubble parameter on the brane. This mode is shown to be 
connected to the excitation of small black hole in the 
bulk in the long wavelength limit. The exact solution for 
an anisotropic stress on the brane induced by this KK mode 
is found, which plays an important role in the calculation
of cosmic microwave background radiation anisotropies in 
the brane world.

\end{abstract}



\maketitle

\section{Introduction}

In the past few years, a lot of efforts have been devoted to the
investigation of the brane world scenario, where our Universe is a
hypersurface, called a brane, embedded in a higher dimensional bulk
spacetime.
Especially, models proposed by Randall and Sundrum have attracted much
attention in the context of gravity and cosmology
\cite{RSI}-\cite{review}.
In their second model (RS model), a positive tension brane is embedded in
five-dimensional anti-de Sitter (AdS) spacetime.
The standard model particles are confined to the brane while the
gravity can propagate in the bulk.
An interesting feature of their model is that four-dimensional gravity
can be recovered at low energy despite the infinite size of the extra
dimension.
It breaks the conventional idea that the extra dimension must be
compact and small.
The extension of the RS model to dilatonic brane
worlds have been intensively investigated
\cite{Lukas}-\cite{Soda}.

When we discuss the gravity on the brane, it
is useful to derive the effective four-dimensional Einstein equation on
the brane firstly developed by Shiromizu, Maeda, and Sasaki
\cite{SMS,MW}.
The effective four-dimensional Einstein equation
includes the term $E_{\mu\nu}$ that is the
electric part of five-dimensional Weyl tensor. 
This term is induced by the gravitational field 
in the bulk and carries the information in the bulk. 
In the RS model with AdS bulk spacetime, $E_{\mu \nu}$ 
tensor can induce "dark radiation" on the brane 
in the homogeneous and isotropic background spacetime. 
It has been realized that the appearance of the 
dark radiation on the brane is related to the existence of 
the black hole in the bulk. 

The dark radiation provides interesting phenomena
in the brane world cosmology. 
First, it modifies the expansion of the background
universe in the same way as an usual radiation
\cite{Ida, Ichiki}.
Secondly, it also gives important effects on cosmological
perturbations on the brane. The cosmological perturbations in brane world
have been actively discussed
\cite{Maartens0}-\cite{KoyamaSoda3} and 
the possible impact on cosmic microwave background (CMB) anisotropies
of the dark radiation perturbation is discussed \cite{Large,Koyama}.
In this paper, we focus our attention to the dark radiation in
cosmological perturbations.

The difficulty in the calculation of dark radiation perturbation
is that it is no longer the radiation fluid once we consider the 
perturbation. This is because $E_{\mu \nu}$ could have a 
non-trivial component of an anisotropic stress. This renders distinguishable
features to the dark radiation perturbation from an usual radiation fluid. 

Because $E_{\mu \nu}$ is determined by the bulk gravitational field, 
it cannot be determined solely by the four-dimensional equations 
on the brane in general. Nevertheless, it is possible to know some features 
of this tensor by using constraint equations on the brane obtained by the
four-dimensional Bianchi identity. In the background spacetime, the 
four-dimensional equations are sufficient to show that $E_{\mu \nu}$ 
induces the radiation fluid on the brane. In order to determine the 
amplitude of the energy density of dark radiation, the information 
in the bulk, that is, the mass of the black hole in the bulk is needed.  
In the case of the perturbations, it is impossible to determine the 
anisotropic component of $E_{\mu \nu}$ only from the four-dimensional 
equations. It is needed to calculate the perturbations in the bulk. 

The attempt to connect the dark radiation perturbation on the brane 
to the bulk perturbations was made in the Ref. \cite{KoyamaSoda3}. 
However, in the RS model, 
it is impossible to find the analytic solutions for the bulk perturbations 
that properly satisfy the junction conditions at the brane. 
Then it is difficult to analyze the precise relation between the 
dark radiation perturbation and the bulk perturbations. 

In this paper, we use a model provided by Koyama and Takahashi 
\cite{KoyamaTakahashi,KoyamaTakahashi2}.
This model is proposed in the context of an inflationary brane model 
induced by a bulk scalar field \cite{KobayashiKoyamaSoda}-\cite{Himemoto4}.
The great advantage of this model is that the five dimensional 
cosmological perturbations can be solved analytically. 
Very recently, Kobayashi and Tanaka introduced a 
$(5+m)$-dimensional vacuum description of this model 
that makes the analysis of cosmological perturbation simple 
\cite{KobayashiTanaka}. They found the complete sets of the 
solutions for bulk perturbations. 
The main purpose of this paper is to clarify the connection between
the dark radiation perturbation and bulk perturbations in the bulk in this
exactly solvable model.

The plan of this paper is as follows.
In Sec.\ref{back}, we briefly review the background spacetime.
We then derive the four-dimensional effective Einstein equations and
the equation of motion for the scalar field on the brane in
Sec.\ref{effective}. In the dilatonic brane world, $E_{\mu\nu}$
contains the contribution from the bulk scalar field. In order to 
make it easy to compare our analysis with the one in the RS model, 
we separate the contribution from the bulk scalar
field in $E_{\mu \nu}$ and define a new tensor $F_{\mu \nu}$
which contains the information of the bulk gravitational fields.
In Sec.\ref{cosmological}, we discuss the "dark radiation" in
cosmological perturbations on the brane.
First, we find the dark radiation like solution  
for the constraint equations for $F_{\mu \nu}$
obtained by the four-dimensional Bianchi identity.
We also calculate the exact solutions for $F_{\mu \nu}$ 
using the solutions of the bulk gravitational field obtained in
\cite{KoyamaTakahashi,KoyamaTakahashi2,KobayashiTanaka}.
Then comparing these two results, it is possible to identify 
the bulk perturbation that induces the dark radiation like 
contributions on the brane. 
In Sec.\ref{dark}, we discuss the connection between this bulk 
perturbation and the excitation of perturbatively small black hole 
in the bulk.  In Sec.\ref{summary}, we summarize the 
results and discuss the anisotropic
stress induced by the dark radiation perturbation and its
implication for the CMB anisotropies. 

\section{BACKGROUND SPACETIME}
\label{back}

We first review the background spacetime
\cite{KoyamaTakahashi,KoyamaTakahashi2}.
We start from the five-dimensional Einstein-Hilbert action with
a bulk scalar field,
%
\begin{eqnarray}
S = \int d^{5}x \sqrt{- g_{5}} \left( \frac{1}{2 \kappa^{2}} R 
- \frac{1}{2} \partial_{\mu} \varphi \partial^{\mu} \varphi - \Lambda(\varphi)
\right) \nonumber \\
- \int d^{4}x \sqrt{- g_{4}} \lambda(\varphi),
\end{eqnarray}
where $\kappa^{2}$ is five-dimensional gravitational constant. 
The potential for the scalar field in the bulk and on the brane are taken to be exponential:

\begin{eqnarray}
\kappa^{2} \Lambda(\varphi) & = & \left( \frac{\Delta}{8} + \delta \right) \lambda^{2}_{0}
e^{-2 \sqrt{2} b \kappa \varphi}, 
\label{bulk_pot}
\\ 
\kappa^{2} \lambda(\varphi) & = & \sqrt{2} \lambda_{0} e^{- \sqrt{2} b
\kappa \varphi}.
\label{brane_pot}
\end{eqnarray}
Here $\lambda_{0}$ is the energy scale of the potential, $b$ is the dilaton coupling 
and we defined 
%
\begin{equation}
\Delta = 4 b^{2} - \frac{8}{3}.
\end{equation}
We assume the $Z_2$ symmetry across the brane.
This type of scalar field arises from a sphere reduction in M theory
or string theory.

For $\delta=0$, the static brane solution was found \cite{Cvetic}. 
The existence of the static brane requires tunning between bulk potential
and brane tension known as Randall-Sundrum tunning.
It has been shown that for $\Delta \leq -2$, we can avoid the presence
of the naked singularity in the bulk and also ensure the trapping of
the gravity.
The reality of the dilaton coupling requires $-8/3 \leq \Delta$.
Thus in the rest of the paper we shall assume $-8/3 < \Delta < -2$.
For $\Delta=8/3$, we recover the Randall-Sundrum model. 
The value of $\delta$ represents a
deviation from the Randall-Sundrum tuning.
This deviation drives an inflation on the brane.

The solution for background spacetime is found as
\begin{eqnarray}
ds^{2} &=& e^{2 W(y)} \left( - dt^{2} + e^{2 \alpha(t)} \delta_{ij} dx^{i} dx^{j}
+ e^{2 \sqrt{2}b \kappa \varphi(t)} dy^{2} \right),\nonumber\\
\varphi(t,y) &=& \varphi(t) + \Xi(y).
\label{back_metric}
\end{eqnarray}
The evolution equations for $\alpha(t)$ and $\varphi(t)$ are given by
\begin{equation}
\dot{\alpha}^2+\sqrt{2}b \kappa \dot{\varphi} \dot{\alpha}=
\frac{1}{6} \kappa^2 \dot{\varphi}^2-\frac{1}{3}\lambda_0^2 \frac{\Delta+4}{\Delta}
\delta e^{-2 \sqrt{2}b \kappa \varphi},
\label{back_alpha}
\end{equation}
\begin{equation}
\ddot{\varphi}+(3 \dot{\alpha} + \sqrt{2} b \kappa \dot{\varphi}) \dot{\varphi}
=-4 \sqrt{2} b \kappa^{-1} \lambda_0^2 \frac{\delta}{\Delta} e^{-2 \sqrt{2}b \kappa \varphi},
\label{back_scalar}
\end{equation}
where dot denotes the derivative with respect to $t$. 

The solution for $\alpha(t)$ and $\varphi(t)$ can be easily found as 
\begin{eqnarray}
e^{\alpha(t)} &=& (H_0 t)^{\frac{2}{3 \Delta+8}}=(- H \eta)^{\frac{2}{3(\Delta+2)}} ,
\label{eq:solution_alpha}
\\
e^{ \sqrt{2} b \kappa \varphi(t)} &=& H_0 t=\left( - H \eta \right)^{\frac{3 \Delta + 8}{3 (\Delta + 2)}}.
\label{eq:solution_phi}
\end{eqnarray}
where 
\begin{equation}
H_0 \equiv -\frac{3\Delta+8}{3 (\Delta +2)} H, \quad 
H=- (\Delta + 2) \sqrt{- \frac{\delta}{\Delta}} \lambda_{0},
\end{equation}
and a conformal time $\eta$ is defined as 
\begin{equation}
\eta = \int e^{-\alpha} dt
= \frac{3 \Delta + 8}{3 (\Delta + 2)} H_{0}^{-\frac{2}{3 \Delta + 8}} 
t^{\frac{3 (\Delta + 2)}{3 \Delta + 8}}.
\end{equation}
We should notice that power-law inflation occurs on the brane 
for $-8/3 < \Delta < -2$.

The solutions for $W(y)$ and $\Xi(y)$ can be written as 
\begin{equation}
e^{W(y)} = {\cal H}(y)^{\frac{2}{3(\Delta+2)}},\quad e^{\kappa \Xi(y)}=
{\cal H}(y)^{\frac{2 \sqrt{2}b}{(\Delta+2)}},
\end{equation}
where
\begin{equation}
{\cal H}(y)= \frac{\sinh{H y}}{\sinh{H y_0}}, \quad 
\sinh Hy_0=\frac{1}{\sqrt{-1-\frac{\Delta}{8 \delta}}}.
\label{eq:background_negative} 
\end{equation}
Here we assumed $\frac{\Delta}{8} + \delta < 0$.

The above five-dimensional solution can be
obtained by a coordinate transformation from the metric
\begin{equation}
ds^2 = e^{2P(z)}(dz^2 -d \tau^2 + \delta_{i j} dx^i dx^j),\quad 
e^{\kappa \varphi(z)}=e^{3 \sqrt{2} b P(z)},
\label{static}
\end{equation}
where
\begin{equation}
e^{P(z)} = (\sinh H y_0)^{-\frac{2}{3(\Delta+2)}} (H
z)^{\frac{2}{3(\Delta+2)}},
\end{equation}
by
\begin{eqnarray}
z = - \eta \sinh (H y), \nonumber\\
\tau = -\eta \cosh (H y). 
\label{trans}
\end{eqnarray}
The metric Eq. (\ref{static}) is often convenient because
of its simplicity.
(In the section \ref{dark}, we calculate the behavior of
five-dimensional Weyl tensor
in the presence of perturbatively small mass of black hole in the bulk
not directly in Eq. (\ref{back_metric}) but in Eq. (\ref{static}).)

The background equations on the brane Eq.(\ref{back_alpha}) and
(\ref{back_scalar}) can be
described by the four-dimensional Brans-Dicke theory with the action
\begin{eqnarray}
S_{4,eff}&=&\frac{1}{2 \kappa_4^2} \int d^4 x \sqrt{-g_4} \left[ 
\varphi_{BD} \: ^{(4)}R-\frac{\omega_{BD}}{\varphi_{BD}}(\partial
\varphi_{BD})^2 \right] 
\nonumber \\&&- \int d^4 x \sqrt{-g_4} V_{eff}(\varphi_{BD}),
\label{eq:BD}
\end{eqnarray}
where 
\begin{eqnarray}
\varphi_{BD}=e^{\sqrt{2}b \kappa \varphi}, \quad 
\omega_{BD}=\frac{1}{2 b^2}, \quad 
\nonumber\\
\kappa_4^2 V_{eff}(\varphi_{BD}) =- \lambda_0^2 \delta
\frac{\Delta+4}{\Delta} \frac{1}{\varphi_{BD}}.
\end{eqnarray}

\section{EFFECTIVE EQUATIONS ON THE BRANE}
\label{effective}

In this section, we derive the effective gravitational equations
on the brane using the covariant curvature formalism developed in
\cite{SMS,MW}.
Using the Gauss equation and the Israel's junction condition, we
obtain the induced four-dimensional Einstein equations on the brane as
%
\begin{eqnarray}
{}^{(4)}G_{\mu\nu}=-{}^{(4)}\Lambda q_{\mu\nu} + \frac{2}{3}\kappa^2
T_{\mu\nu}^{(b)}-E_{\mu\nu},
\label{4Deqn}
\end{eqnarray}
%
where
%
\begin{eqnarray}
T^{(b)}_{\mu\nu}=D_\mu \varphi D_\nu \varphi - \frac{5}{8} q_{\mu\nu}
(D\varphi)^2,
\label{Tdef}
\end{eqnarray}
%
%
\begin{eqnarray}
{}^{(4)}\Lambda &=& \frac{\kappa^2}{2}\left[ \Lambda +
\frac{\kappa^2}{6}\lambda^2 -\frac{1}{8} \left(
\frac{d\lambda}{d\varphi}  \right)^2 
\right]
\nonumber\\
&=&\frac{\delta}{2} \lambda_0^2 e^{-2\sqrt{2}b\kappa \varphi},
\label{4dcos_con}
\end{eqnarray}
%
and
%
\begin{eqnarray}
E_{\mu\nu} = {}^{(5)}C_{\mu\alpha\nu\beta} n^{\alpha} n^{\beta}.
\end{eqnarray}
%
We defined $D_\mu$ as the covariant derivative with respect
to the induced metric on the brane.
We note that four-dimensional cosmological constant ${}^{(4)} \Lambda$
is proportional to $\delta$, which represents a deviation from the
Randall-Sundrum tuning.
The 4-dimensional gravitational equation on the brane Eq (\ref{4Deqn})
includes the projected Weyl tensor $E_{\mu\nu}$, which can not be
determined without solving the bulk dynamics in general.
This term plays an essential role when we consider the cosmological
perturbations in brane world models.
%
%
Taking the divergence of the four-dimensional effective equations and 
using four-dimensional Bianchi
identity, we obtain the constraint equations for $E_{\mu\nu}$ as
%
\begin{eqnarray}
D^{\mu}E_{\mu\nu} = \frac{2\kappa^2}{3} D^{\mu}T^{(b)}_{\mu\nu}
-D^{\mu} {}^{(4)}\Lambda q_{\mu\nu}.
\label{conE2}
\end{eqnarray}

Because $E_{\mu \nu}$ contains the contribution from the bulk scalar
field, it is convenient to separate the contributions of 
the bulk scalar field from $E_{\mu\nu}$. We define
%
\begin{eqnarray}
-E_{\mu\nu} &=& 
\sqrt{2}b\kappa \left(D_\mu D_\nu \varphi-q_{\mu\nu} D^2 \varphi\right) 
\nonumber\\
&&+ 2b^2\kappa^2 \left(D_\mu \varphi D_\nu \varphi - q_{\mu\nu}
(D\varphi)^2 \right)
\nonumber\\
&&+ \frac{\kappa^2}{3} \left(D_\mu \varphi D_\nu \varphi - \frac{1}{4}
q_{\mu\nu} (D\varphi)^2 \right)
\nonumber\\
&&+\frac{6b^2}{\Delta} \lambda_0^2 \delta e^{-2\sqrt{2}b\kappa\varphi}
q_{\mu\nu}
+F_{\mu\nu},
\label{defF}
\end{eqnarray}
%
We also rewrite the the equation of motion for the scalar field on the brane 
as 
%
\begin{eqnarray}
D^2 \varphi + \sqrt{2}b\kappa (D\varphi)^2 - 4\sqrt{2}
\frac{b}{\kappa} \lambda_0^2 \frac{\delta}{\Delta}
e^{-2\sqrt{2}b\kappa \varphi} = F_{\varphi}.
\label{eomF}
\end{eqnarray}
%
From traceless condition of $E_{\mu\nu}$, $F^\mu_\mu$ are related to
$F_\varphi$ as
%
\begin{eqnarray}
F^\mu_\mu=3\sqrt{2}b\kappa F_{\varphi}.
\label{traceF}
\end{eqnarray}
%
The equations derived from the effective action Eq.(\ref{eq:BD})
agree with Eqs. (\ref{defF}), (\ref{eomF}) with $F_{\mu \nu}=0$
and $F_{\varphi}=0$. Thus $F_{\mu \nu}$ and $F_{\varphi}$
are expected to describe the contribution of KK modes.
It should be noted that a similar decomposition of $E_{\mu \nu}$
was considered in Ref. \cite{Minamitsuji2}. 

Substituting the expression for $E_{\mu\nu}$ Eq.(\ref{defF}) and
using the equation of motion for the scalar field Eq.(\ref{eomF}), 
we can rewrite the 4D Bianchi identity Eq.(\ref{conE2}) as 
%
\begin{eqnarray}
D^{\mu} F_{\mu\nu} + \sqrt{2}b\kappa D^{\mu} \varphi F_{\mu\nu} =
-\kappa^2D_{\nu} \varphi  F_{\varphi}.
\label{relationF}
\end{eqnarray}
%
In general, this constraint equation is not sufficient to completely
determine the behavior of $F_{\mu\nu}$ and $F_{\varphi}$ on the brane.

\section{DARK RADIATION IN COSMOLOGICAL PERTURBATIONS}
\label{cosmological}

In this section, we consider the dark radiation in cosmological
perturbations on the brane.
Since the dark radiation corresponds to scalar type perturbations, we
restrict our attention to cosmological perturbations of this type
throughout the paper.
It is assumed $F_{\mu\nu}=0$ and $F_\varphi=0$ for the background
spacetime in Koyama-Takahashi model.

First, we show that the dark radiation appears as a solution of the
constraint equations for $\delta F_{\mu\nu}$ and $\delta F_\varphi$ on
the brane Eq.(\ref{relationF}) at large scales in the section
\ref{viewbrane}.
$\delta F_{\mu\nu}$ and $\delta F_\varphi$ have four
independent variables for scalar perturbations.
We also show that two of the four variables can not be determined by
their constraint equations Eq.(\ref{relationF}).

Next, we calculate the exact solution of $\delta F_{\mu\nu}$ and
$\delta F_\varphi$ using the solutions for the five-dimensional
perturbed Einstein equations obtained by
\cite{KoyamaTakahashi,KoyamaTakahashi2,KobayashiTanaka}
in the section \ref{viewbulk}.
We then investigate the relation between the dark radiation and 
the bulk perturbations.

\subsection{View from the brane}
\label{viewbrane}

Here we consider the dark radiation as a solution of the
constraint equations for $\delta F_{\mu\nu}$ and $\delta F_\varphi$ on
the brane Eq.(\ref{relationF}).
First of all, we expand $\delta F_{\mu\nu}$ in terms of the scalar
harmonics as
%
\begin{eqnarray}
&&\delta F_{tt} = \delta \rho_F Y,
\nonumber\\
&&\delta F_{ti} = e^{\alpha} \delta q_F Y_{i},
\nonumber\\
&&\delta F_{ij} = e^{2\alpha} \biggl( \frac{1}{3} (\delta \rho_F +
3\sqrt{2}b\kappa \delta F_\varphi ) Y \delta_{ij}
\nonumber\\&&~~~~
+ \delta \pi_F Y_{ij} \biggr),
\end{eqnarray}
%
where $Y(k,x) \propto e^{i k x}$ is the normalized scalar harmonics and 
the vector $Y_i$ and traceless tensor $Y_{ij}$ are constructed 
from $Y$ as $Y_i=- k^{-1} Y_{,i} , \quad Y_{ij}= k^{-2} Y_{,ij}
+\delta_{ij} Y/3$.

The four-dimensional perturbed Einstein equations and the equation for
the scalar field are not closed but include four variables
$\delta \rho_F, \delta q_F, \delta \pi_F$ and $\delta F_\varphi$. 
The concrete forms of these equations are presented in Appendix A.
On the other hand, there are constraint equations on $\delta
F_{\mu\nu}$ and $\delta F_\varphi$, Eq.(\ref{relationF}).
For scalar type perturbations, these become two equations as
follows:
%
\begin{eqnarray}
( \partial_t + 4\dot{\alpha} + \sqrt{2}b\kappa \dot{\varphi}) \delta
 \rho_F - ke^{-\alpha} \delta q_F = 0,
\label{relationFtp}
\end{eqnarray}
%
%
\begin{eqnarray}
&&( \partial_t + 4\dot{\alpha} + \sqrt{2}b\kappa \dot{\varphi}) \delta
 q_F 
\nonumber\\&&~~~~
+ ke^{-\alpha} \biggl(\frac{2}{3} \delta \pi_F - \frac{1}{3}
 \delta  \rho_F - \sqrt{2}b\kappa \delta F_\varphi \biggr) = 0.
\label{relationFip}
\end{eqnarray}
%
Here we used the condition Eq.(\ref{traceF}).
At large scales $k e^{-\alpha}/H \to 0$, we can neglect the term
proportional to $\delta q_F$ in Eq.(\ref{relationFtp}).
The solution of $\delta \rho_F$ is given by $\delta \rho_F = \delta C
e^{-4\alpha -\sqrt{2}b\kappa \varphi}$.
This corresponds to the dark radiation (hereafter we call this 
dark radiation although this does not behave as a radiation 
for $b \neq 0$).
It can be checked that the integration constant $\delta C$ is related
to the perturbatively small black hole mass in the bulk (see the
section \ref{dark}).
On the other hand, we can not determine $\delta \pi_F$ because it is
dropped from Eq.(\ref{relationFip}) for $k e^{-\alpha}/H \to 0$.
This uncertainty prevents us from predicting CMB anisotropies in brane
world models \cite{Large,Koyama}.
This issue is discussed in the section \ref{summary}.

Since there are only two constraint equations, two of the four
variables can not be determined.
On the other hand, there are two physical degrees of freedom in the
bulk for scalar type perturbations.
One of them corresponds to the scalar field and the other to the
graviscalar.
To investigate the relation between the dark radiation and the bulk
perturbations, 
we need to obtain the exact solutions for $\delta F_{\mu\nu}$ and
$\delta F_\varphi$.
This can be achieved only when we solve the bulk gravitational field
and determine the behavior of the two unknown variables.

\subsection{View from the bulk}
\label{viewbulk}

In the previous subsection, we showed that we must solve the bulk
gravitational field in order to completely determine the 
contributions of $F_{\mu \nu}$ and $F_{\varphi}$ to cosmological 
perturbations on the brane. Here, we first summarize the two independent 
solutions of five-dimensional Einstein equations for scalar perturbations 
obtained in \cite{KobayashiTanaka} in Sec.\ref{solutions}.
Using these solutions, we then calculate and investigate the behavior
of the contributions of $F_{\mu \nu}$ and $F_{\varphi}$ 
to cosmological perturbations on the brane in Sec.\ref{kk}.
Finally we discuss their relation to the dark radiation in
Sec.\ref{kkmass}.


\subsubsection{Solutions of the bulk gravitational field}
\label{solutions}

Here, we present the two independent solutions of
five-dimensional Einstein equations for scalar perturbations obtained
in \cite{KobayashiTanaka}.
The perturbed metric and scalar field are given by
%
\begin{eqnarray}
ds^2 &=& e^{2 W(y)} \biggl[ e^{2 \sqrt{2}b \kappa \varphi(t)}(1+2NY)dy^2 
+2 AY  dt dy 
\nonumber\\&&
-(1+2 \Phi Y )dt^2+ 
e^{2 \alpha(t)} \biggl((1+2 \Psi Y)\delta_{ij} dx^i dx^j 
\nonumber\\&&
+ 2E Y_{ij} dx^i dx^j + 2B Y_i dx^i dt + 2C Y_i dx^i dy \biggr) \biggr], 
\nonumber\\
\varphi &=& \varphi(t)+\Xi(y)+\delta \varphi Y.
\end{eqnarray}
%
We note that any gauge condition is not imposed here.

Under a scalar gauge transformation,
%
\begin{eqnarray}
t &\to& \bar{t}=t+\xi^t Y,
\nonumber\\
y &\to& \bar{y}=y+\xi^y Y,
\nonumber\\
x^i &\to& \bar{x}^i=x^i+\xi^S Y^i,
\end{eqnarray}
%
the metric variables transform as
%
\begin{eqnarray}
N &\to& \bar{N}=N-\xi^y {}' -W'\xi^y-\sqrt{2}b\kappa\dot{\varphi}\xi^t,
\nonumber\\
A &\to& \bar{A}=A+\xi^t {}'-e^{2\sqrt{2}b\kappa\varphi}\dot{\xi}^y,
\nonumber\\
C &\to& \bar{C}=C+e^{-2\alpha+2\sqrt{2}b\kappa\varphi}k\xi^y -\xi^S{}',
\nonumber\\
\Phi &\to& \bar{\Phi}=\Phi-W'\xi^y -\dot{\xi}^t,
\nonumber\\
B &\to& \bar{B}=B-e^{-2\alpha}k\xi^t -\dot{\xi}^S,
\nonumber\\
E &\to& \bar{E}=E+k\xi^S,
\nonumber\\
\Psi &\to& \bar{\Psi}=\Psi-\frac{1}{3}k\xi^S -W'\xi^y
-\dot{\alpha}\xi^t,
\nonumber\\
\delta \varphi &\to& \delta \bar{\varphi}=\delta \varphi
-\dot{\varphi}\xi^t -\Xi'\xi^y,
\label{gauge}
\end{eqnarray}
%
where prime denotes the derivative with respect to $y$.
In our background spacetime, the gauge fixing condition imposed in
\cite{KobayashiTanaka} corresponds to
%
\begin{eqnarray}
N=\sqrt{2}b\kappa \delta \varphi, ~~A=C=0.
\label{gauge_cond}
\end{eqnarray}
%
These conditions do not fix the gauge completely.
We use this remaining degree of freedom to keep the brane location
unperturbed at $y=y_0$.
Then, the boundary conditions on the brane for all the
remaining variables become Neumann boundary condition;
%
\begin{eqnarray}
\partial_y N |_{y_0} = \partial_y \Phi |_{y_0} = \partial_y \Psi
|_{y_0} = \partial_y E |_{y_0} = \partial_y B |_{y_0} =0.
\end{eqnarray}
%

All variables can be expanded by the same mode functions in the
$y$-direction as
%
\begin{eqnarray}
\Phi = \Phi_0 (t) \psi_0 (y) + \sum \Phi_m (t) \psi_m (y), ...,
\label{solutiony}
\end{eqnarray}
%
where $\psi_0$ is constant, and
%
\begin{eqnarray}
&&\psi_m (y) = c ({\rm sinh} H y)^{1/2 + \mu} 
B^{-1/2-\mu}_{-1/2+i\nu}( {\rm cosh} Hy),\nonumber\\~~~&&
B^{\beta}_{\gamma}({\rm cosh} Hy)
=Q^{1/2-\mu}_{-1/2+i\nu} ({\rm cosh} H y_0)
P^{\beta}_{\gamma}  ({\rm cosh} H y)
\nonumber\\~~~&&
- P^{1/2-\mu}_{-1/2+i\nu} ({\rm cosh} H y_0)
Q^{\beta}_{\gamma}  ({\rm cosh} H y),
\label{modey}
\\
&&\mu=-\frac{1}{(\Delta+2)},
\\
&&\nu(m) = \sqrt{\frac{m^2}{H^2} - \mu^2},
\end{eqnarray}
%
where $c$ is a normalization constant.
The first and second terms in Eq.(\ref{solutiony}) represent
the zero and KK modes, respectively.
$m^2$ represents the squared KK mass for observers on the
four-dimensional brane.
There is a mass gap $\delta m = \mu H$ between the zero mode and the
KK continuum.
The modes with $0<m<\mu H$ are not normalizable.
$P^\alpha_\beta$ and $Q^\alpha_\beta$ are associated
Legendre functions.

We now turn to the mode functions in the $t$-direction.
For the zero mode, we have another gauge degrees of freedom.
As is evident from Eq.(\ref{gauge}), gauge transformation satisfying
$\xi^y=0$ and $\xi^t {}'=\xi^S {}'=0$ do not disturb the conditions
Eq.(\ref{gauge_cond}).
We can use this degree of freedom to set $B=E=0$, because the
solutions do not depend on $y$ for the zero mode.
The solutions are given by
%
\begin{eqnarray}
N_0 &=& \sqrt{2}b\kappa \delta \varphi_0
\nonumber\\
&=& c_0 \frac{1}{3} \frac{\Delta+2}{\Delta+3} \biggl(\rho_\mu -
\frac{3\Delta+8}{\Delta+4} \rho_{\mu-2}\biggr),
\nonumber\\
\Psi_0 &=& - c_0 \frac{2}{3} \frac{\Delta+2}{\Delta+3} \biggl(\rho_\mu +
\frac{1}{\Delta+4} \rho_{\mu-2}\biggr),
\nonumber\\
\Phi_0 &=& -\Psi_0 - N_0,
\end{eqnarray}
%
where $c_0$ is a constant and $\rho_\alpha$ is defined as
%
\begin{eqnarray}
\rho_\alpha (\eta) = (- k \eta)^\mu H_\alpha (-k\eta),
\label{modet}
\end{eqnarray}
%
where $H_\alpha$ is an arbitrary linear combination of Hankel
functions $H^{(1)}_\alpha$ and $H^{(2)}_\alpha$.
We note that the number of physical degree of freedom is one for the
zero mode.
This solution is already obtained by Koyama and Takahashi
\cite{KoyamaTakahashi,KoyamaTakahashi2}.

The solutions for the KK modes in the gauge condition
Eq.(\ref{gauge_cond}) are obtained as
%
\begin{eqnarray}
N_m &=& \sqrt{2}b\kappa \delta \varphi_m
\nonumber\\
&=& - \frac{1}{3} \frac{3\Delta+8}{\Delta+4} \biggl[c_1 
\biggl((2\mu-1)k\eta\rho_{i\nu-1}
\nonumber\\&&
+ (i\nu+\mu)(i\nu+\mu-1) \rho_{i\nu} \biggr)
+ 2 c_2 \rho_{i\nu} \biggr],
\\
\Phi_m &=& c_1 (-k \eta)^2 \rho_{i\nu} + N_m,
\\
\Psi_m &=& -\frac{1}{3} c_1 (- k \eta)^2 \rho_{i\nu},
\\
E_m &=& c_1 (-k \eta)^2 \rho_{i\nu} - \frac{2}{\Delta+4} c_1 
\biggl((2\mu-1)k\eta\rho_{i\nu-1} 
\nonumber\\&&
+ (i\nu+\mu)(i\nu+\mu-1) \rho_{i\nu} \biggr)
+ \frac{3\Delta+8}{\Delta+4} c_2 \rho_{i\nu},
\\
B_m &=& 2 c_1 e^{-\alpha} k \eta \biggl((i\nu+\mu-1)\rho_{i\nu} +
k\eta\rho_{i\nu-1}\biggr).
\end{eqnarray}
%
We should note that the solution obtained in 
\cite{KoyamaTakahashi2} is a particular solution where $c_1$ and $c_2$
are related (see Appendix C). 

\subsubsection{Solutions for $F_{\mu \nu}$ and $F_{\varphi}$}
\label{kk}

Next, we calculate the solutions for $\delta
F_{\mu\nu}$ and $\delta F_\varphi$, using the solutions of the bulk
gravitational field summarized above.
The above two solutions are obtained in the Gaussian-normal gauge
condition with respect to the brane.
After a gauge transformation to the longitudinal gauge (see Appendix
C), we substitute the solutions projected on the brane into the
four-dimensional perturbed Einstein equations in Appendix A,
Eq.(\ref{4Dtt_bianchi}), (\ref{4Dij_bianchi}), (\ref{4Dti_bianchi}),
and (\ref{eqQ}).
Then we obtain $\delta F_{\mu\nu}$ and $\delta F_\varphi$ as
%
\begin{eqnarray}
&&\delta \rho_F = - \tilde{c}_1 e^{-2\alpha} \rho_{i\nu} (\eta) \psi_m
(y_0),
\\
&&\delta q_F = \tilde{c}_1 e^{-2\alpha} (k\eta)^{-1} \biggl( (i\nu +
\mu -1 )\rho_{i\nu} (\eta) 
\nonumber\\&&~~
+ k\eta \rho_{i\nu-1} (\eta) \biggr)\psi_m
(y_0),
\\
&&\delta \pi_F = \tilde{c}_1 (i\nu + \mu -1) e^{-2\alpha} \biggl[
\frac{1}{i\nu-1} \biggl( \rho_{i\nu} (\eta) 
\nonumber\\&&~~
+ \frac{\mu}{i\nu+\mu-1} \rho_{i\nu-2} (\eta) \biggr)
-2 \frac{i\nu+\mu}{\Delta+4} \frac{\rho_{i\nu} (\eta)}{k^2\eta^2}
\biggr] \psi_m(y_0)
\nonumber\\&&~~
+ \tilde{c}_2 \frac{3\Delta+8}{\Delta+4} e^{-2\alpha} (k \eta)^{-2}
\rho_{i\nu} (\eta) \psi_m (y_0),
\\
&& \delta F_\varphi = \tilde{c}_1 \sqrt{2}b\kappa^{-1} (i\nu + \mu -1)
e^{-2\alpha} \biggl[ 2 \frac{i\nu+\mu}{\Delta+4} \frac{\rho_{i\nu}
(\eta)}{k^2\eta^2}
\nonumber\\&&~~
-\frac{\mu}{(i\nu + \mu -1) (i\nu-1)} 
( \rho_{i\nu} (\eta) + \rho_{i\nu-2} (\eta) )
\biggr] \psi_m(y_0)
\nonumber\\&&~~
+ \tilde{c}_2 \frac{4\sqrt{2}b\kappa^{-1}}{\Delta+4} e^{-2\alpha}
(k\eta)^{-2} \rho_{i\nu} (\eta) \psi_m (y_0),
\end{eqnarray}
%
where
%
\begin{eqnarray}
&&\tilde{c}_1=(i\nu+\mu)(i\nu-\mu) k^2 c_1 = -\frac{m^2}{H^2} k^2 c_1,
\\
&&\tilde{c}_2=(i\nu+\mu)(i\nu-\mu) k^2 c_2 = -\frac{m^2}{H^2} k^2 c_2.
\end{eqnarray}
%
As is expected, $\delta F_{\mu\nu}$ and $\delta F_\varphi$ vanish for
$m^2=0$.

%
%

\subsubsection{Dark radiation and bulk perturbation}
\label{kkmass}

In Sec.\ref{kk}, we calculated the contributions of the bulk 
perturbation to $F_{\mu \nu}$ and $F_{\varphi}$ on the brane.
Here, we discuss their relation to the dark radiation.
As mentioned before, there are two physical degrees of freedom in the
bulk for scalar perturbations.
One of them corresponds to the scalar field and the other to the
graviscalar.
Since the $c_2$ component of $\delta \rho_F$ vanishes, it is expected
that the solution of $c_1$ includes the dark radiation at large
scales and thus corresponds to the graviscalar.
This can be explicitly shown if we take 
%
$i\nu + \mu -1 = 0$ 
%
and a linear combination of Hankel functions:
%
\begin{eqnarray}
H^{(1)}_\alpha(-k\eta) + e^{-2i\alpha \pi} H^{(2)}_\alpha(-k\eta) = 2
e^{-i\alpha \pi} J_{-\alpha},
\end{eqnarray}
%
such that $H_\alpha (-k\eta) \propto (-k\eta)^{-\alpha}$ for $-k\eta
\to 0$.
In this case, the above solutions for $\delta F_{\mu\nu}$ and $\delta
F_\varphi$ becomes
%
\begin{eqnarray}
\delta \rho_F &=& -(1-2\mu) c_1 e^{-2\alpha} \rho_{1-\mu} \psi_m (y_0)
\nonumber\\
&\propto& - \tilde{c}_1 e^{-4\alpha-\sqrt{2}b\kappa\varphi},
\\
\delta q_F &\propto& - \tilde{c}_1 \frac{-k\eta}{2\mu}
e^{-4\alpha-\sqrt{2}b\kappa\varphi},
\\
\delta \pi_F &\propto& - \tilde{c}_1 \frac{(-k\eta)^2}{4\mu(\mu+1)}
e^{-4\alpha-\sqrt{2}b\kappa\varphi}
\nonumber\\&&
+ \tilde{c}_2 \frac{3\Delta+8}{\Delta+4} \eta^{-2}
e^{-4\alpha-\sqrt{2}b\kappa\varphi},
\label{stressF}
\\
\delta F_\varphi &\propto& \tilde{c}_1 \sqrt{2}b\kappa^{-1}
e^{-4\alpha-\sqrt{2}b\kappa\varphi}
\nonumber\\&&
+ \tilde{c}_2 \frac{4\sqrt{2}b\kappa^{-1}}{\Delta+4} \eta^{-2}
e^{-4\alpha-\sqrt{2}b\kappa\varphi},
\end{eqnarray}
%
for $-k\eta \to 0$.
The time dependence of $\delta \rho_F$ coincides with that of the
solution of the constraint equation Eq.(\ref{relationFtp}) for large
scales.
It should be noted that the condition $i\nu+\mu-1=0$ can be written as
%
\begin{eqnarray}
m^2= (2\mu-1) H^2 (< \mu^2 H^2).
\end{eqnarray}
%
This is quite an interesting result.
The dark radiation corresponds to a non-normalizable KK mode.
For RS model $b=0$, the mass squared become $2H^2$. 

It should be also emphasized that the behavior of $\delta \pi_F$ which
corresponds to the dark radiation is obtained here
(Eq.(\ref{stressF})).
This variable can not be known by the constraint equation
Eq.(\ref{relationFip}) because $\delta \pi_F$ is dropped for $k \to
0$.
As mentioned above, this uncertainty prevents us from predicting CMB
anisotropies in brane world models \cite{Large,Koyama}.
This issue is discussed in the section \ref{summary}.

\section{BLACK HOLE IN THE BULK AND A KK MODE}
\label{dark}

In the previous section, we showed that the dark radiation corresponds
to a non-normalizable KK mode of cosmological perturbations.
Here, we discuss the connection between this KK mode and
the black hole in the bulk when the black hole mass is perturbatively
small.

There is a black hole solution with a bulk scalar field with an
exponential potential, which coincides with the background spacetime
of the Koyama-Takahashi model when the black hole mass vanishes.
We first review this black hole solution in the section \ref{black}.
We also calculate the behavior of $E_{\mu\nu}$ in the case where
the black hole mass is perturbatively small.
On the other hand, we calculate the perturbation of $E_{\mu\nu}$ using
the solutions of the perturbed five-dimensional Einstein equations for
$i\nu+\mu-1=0$.
It is shown that the asymptotic behavior of $E_{tt}$ in the bulk
coincides with the one which originate from the black
hole in the bulk.

%
%

\subsection{Black hole solution with a bulk scalar field}
\label{black}

Here we review a black hole solution with a bulk scalar field which
has the exponential potential in the bulk Eq.(\ref{bulk_pot}).
When the black hole mass vanishes, this solution coincides with the
background spacetime of Koyama-Takahashi model.

We can find a static solution for the bulk with vanishing cosmological
constant as \cite{Sasaki,LM,CaiJiSoh}
%
\begin{eqnarray}
ds^2 &=& -h(R)dT^2 + \frac{R^{3\Delta+8}}{h(R)} dR^2 + R^2 \delta_{ij}
dx^i dx^j,
\\
\varphi &=& 3\sqrt{2}b\kappa^{-1} {\rm ln}(R)
\end{eqnarray}
%
where
%
\begin{eqnarray}
h(R)={\tilde\lambda}_0^2 R^2 - C R^{6b^2-2},
\end{eqnarray}
%
$C$ is an arbitrary constant and related to black hole mass.
Here we defined
%
\begin{eqnarray}
{\tilde\lambda}_0^2 = \frac{\lambda_0^2}{18} \left(1+\delta
\frac{8}{\Delta} \right).
\end{eqnarray}
%
For $b=0$, this solution becomes AdS-Schwartzshild.
The Friedmann equation on the brane with the tension
Eq.(\ref{brane_pot}) is obtained as \cite{Sasaki}
%
\begin{eqnarray}
\left( \frac{\dot{R}}{R} \right)^2 = \frac{4}{9} \left(\frac{\delta}{-\Delta} \right)
\lambda_0^2 R^{-(3\Delta+8)} + C R^{-(3\Delta+16)/2},
\label{Friedmann}
\end{eqnarray}
%
where dot is the derivative with respect to cosmic time on the brane.

For $C=0$, the background spacetime Eq.(\ref{static}) can be
obtained from this metric by a coordinate transformation,
%
%
%
\begin{eqnarray}
R &=& \left(- \frac{3}{2}{\tilde \lambda}_0 (\Delta+2) z
\right)^{2/3(\Delta+2)},
\\
T &=& \frac{\tau}{ {\tilde \lambda}_0 }.
\end{eqnarray}
%
When there is a perturbatively small mass of black hole in the bulk,
the metric Eq.(\ref{static}) is modified as
\begin{eqnarray}
ds^2 = e^{2P(z)}((1+\delta C f(z))dz^2 - (1-\delta C f(z)) d \tau^2 
\nonumber\\+ \delta_{i j} dx^i dx^j),
\label{static_bh}
\end{eqnarray}
where
\begin{eqnarray}
f(z)=\frac{1}{{\tilde\lambda}_0^2} R(z)^{6b^2-4},
\end{eqnarray}
and $\delta C$ is perturbed black hole mass.
It is noted that this modification can not be regarded as a perturbation
at $z\to \infty$ or $R\to 0$.
We focus our attention to the region sufficiently far from the black
hole so that the above modification can be treated as perturbation.
We can easily calculate five-dimensional Weyl tensor in this
coordinate system as
%
\begin{eqnarray}
C_{\tau z\tau z} = - \frac{1}{4} \delta C R(z)^2 ~\partial_z^2 f(z).
\end{eqnarray}
%
By a coordinate transformation Eq.(\ref{trans}), this is related to
Weyl tensor in our background spacetime Eq.(\ref{back_metric}) as
%
\begin{eqnarray}
C_{\eta y\eta y} = \eta^2 H^2 C_{\tau z\tau z}.
\end{eqnarray}
%
Finally, we obtain the behavior of $E_{tt}$:
%
\begin{eqnarray}
\delta E_{tt} = \frac{9}{8} \Delta \delta C
e^{-4\alpha-\sqrt{2}b\kappa\varphi}
\left(\frac{{\rm sinh} Hy}{{\rm sinh} Hy_0} \right)^{2\mu-1}.
\label{ett_bh}
\end{eqnarray}
%
Another components can be obtained by the homogeneity and isotropy of
three-dimensional spatial coordinates and the condition
$E^\mu_\mu=0$.

\subsection{Perturbations of $E_{\mu\nu}$ in the bulk}
\label{perturbations}

Here we calculate perturbations of $E_{\mu\nu}$ in the bulk, using the
solutions of the five-dimensional perturbed Einstein equations for
$i\nu+\mu-1=0$ at large scales.
In Appendix D, we present the perturbation formula for
five-dimensional Weyl tensor in the Gaussian normal gauge.
Substituting the solutions of the bulk gravitational field, we get
%
\begin{equation}
\delta E_{tt} = -\frac{\Delta}{4} c_1 k^2 e^{-2\alpha} \rho_{1-\mu} \left(
\frac{1}{H^2} \partial_y^2 \psi_m(y) - \psi_m(y) \right).
\end{equation}
%
Here we took $i\nu+\mu-1=0$ and $-k\eta\to0$.
Using the solution for $\psi_m(y)$, the y-dependence of $E_{tt}$ 
can be evaluated as
%
\begin{eqnarray}
&&\frac{1}{H^2} \partial_y^2 \psi_m(y) - \psi_m(y) \nonumber\\
&=& -\frac{2 \mu}{1-2 \mu} 
(\sinh Hy)^{1/2+\mu} 
B^{3/2-\mu}_{1/2-\mu}(\cosh Hy).
\end{eqnarray}
%
On the brane, thanks to the junction condition, 
we can show that
%
\begin{equation}
B^{3/2-\mu}_{1/2-\mu}(\cosh Hy_0)=(1-2 \mu)
B^{-1/2-\mu}_{1/2-\mu}(\cosh Hy_0).
\end{equation}
%
Then $E_{tt}$ on the brane is given by
%
\begin{equation}
\delta E_{tt}(y_0)=
 \frac{\Delta}{2} c_1 k^2 e^{-2\alpha} \mu\rho_{1-\mu} \psi_m (y_0).
\label{ett_c1}
\end{equation}
%
Comparing this solution with Eq.(\ref{ett_bh}),
it is possible to express $c_1$ by the black hole mass $\delta C$ as
%
\begin{eqnarray}
c_1 k^2 = \frac{9}{4} \frac{(-H\eta)^{2\mu-1}}{\mu \rho_{1-\mu}} \psi_m
(y_0)^{-1} \delta C.
\label{c1_C}
\end{eqnarray}
%
We note that the left hand side does not depend on time for $-k\eta
\to 0$.
We can also rewrite the induced four-dimensional Einstein equation
Eq.(\ref{4Deqn}) in accordance with the Friedmann equation
Eq.(\ref{Friedmann}) as
%
\begin{eqnarray}
{}^{(4)}G_{\mu\nu}=\frac{8}{3\Delta} {}^{(4)}\Lambda q_{\mu\nu}
-{\tilde E}_{\mu\nu},
\label{4Deqn_modified}
\end{eqnarray}
%
where
%
\begin{eqnarray}
-{\tilde E}_{\mu\nu} = - E_{\mu\nu} + \frac{2}{3}\kappa^2
 T_{\mu\nu}^{(b)} -\frac{2b^2}{\Delta} \lambda_0^2 \delta
 e^{-2\sqrt{2}b\kappa \varphi} q_{\mu\nu}.
\end{eqnarray}
%
If we substitute the solutions for $i\nu+\mu-1=0$ and then use the
relation Eq.(\ref{c1_C}), ${\tilde E}_{tt}$ becomes
%
\begin{equation}
-\delta {\tilde E}_{tt} = 3\delta C e^{-4\alpha-\sqrt{2}b\kappa
 \varphi}.
\end{equation}
%
Clearly, this corresponds to the dark radiation term in
Eq.(\ref{Friedmann}).
Finally we investigate the y-dependence of the $E_{tt}$.
For large $Hy$, the y-dependence of $E_{tt}$ behaves as
%
\begin{equation}
\frac{1}{H^2} \partial_y^2 \psi_m(y) - \psi_m(y) 
\propto (\sinh Hy)^{2 \mu -1}.
\end{equation}
%
This behavior is precisely the same as $\delta E_{tt}$ 
derived from the BH solution. Thus the correspondence is 
held also in the bulk. 

In the perturbation solutions, there is also anisotropic part of
$E_{\mu\nu}$.
The result for $i\nu+\mu-1=0$ is
%
\begin{equation}
\delta \pi_E  
= - \frac{8}{3\Delta} \frac{(-k\eta)^2}{4\mu(\mu+1)} \delta E_{tt},
\end{equation}
%
where $\delta \pi_E$ is defined in the same way with $\delta
\pi_F$.


\section{SUMMARY AND DISCUSSION}
\label{summary}

In this paper, we discussed the connection between the dark radiation
and the bulk perturbation in a dilatonic brane world 
based a model proposed by Koyama and Takahashi
\cite{KoyamaTakahashi,KoyamaTakahashi2}.

We first derived the four-dimensional effective Einstein equations on
the brane developed in Ref.\cite{SMS,MW}.
We separated the contributions of the bulk scalar field from 
$E_{\mu\nu}$. Then the four-dimensional effective theory becomes 
the BD theory with the corrections given by $F_{\mu\nu}$ and 
$F_\varphi$. We then considered the dark radiation in cosmological 
perturbation on the brane.
The perturbed Einstein equations includes the four variables $\delta
\rho_F, \delta q_F, \delta \pi_F$ and $\delta F_\varphi$ which
carry the information in the bulk.
There are two constraint equations obtained from the four-dimensional 
Bianchi identity. We showed that the dark radiation appears as a 
solution for the constraint equations at large scales.

We can derive a complete set of the solutions for 
$\delta\rho_F, \delta q_F, \delta \pi_F$ and $\delta F_\varphi$ 
only when we solve the bulk gravitational fields, 
which have two physical degrees of freedom, the scalar field
perturbation and the graviscalar. 
We calculated $\delta \rho_F, \delta q_F, \delta \pi_F$ and $\delta
F_\varphi$ on the brane using these two independent solutions of the
bulk perturbations obtained in
Ref.\cite{KoyamaTakahashi,KoyamaTakahashi2,KobayashiTanaka}.
We found that if we take a non-normalizable KK mode with mass
$m^2=(2\mu-1) H^2$, the contribution from the graviscalar in the
bulk corresponds to the dark radiation at large scales.
We also checked that this solution corresponds to 
the excitation of a small black hole in the bulk 
by calculating $\delta E_{tt}$. 
It was shown that the asymptotic behavior of $\delta E_{tt}$ 
induced by the KK mode with mass $m^2=(2\mu-1) H^2$ precisely agrees with
the one derived from the black hole solution with a small black hole mass.

The perturbation of the small black hole mass breaks down 
as approaching to the black hole. We need to include the non-linear 
effect. Thus we expect that un-normalizability of the mode is cured 
if we take into account the non-linear effect. 
For the homogeneous and isotropic case, we know the resultant 
non-linear solution, that is the black hole spacetime. 
A nontrivial result here is that the KK mode also induces an anisotropic
component of $E_{\mu \nu}$ which vanishes in the long wavelength limit. 
This anisotropic component has the same y-dependence as $\delta E_{tt}$. 
Thus in order to discuss the bulk geometry at large $H y$ with anisotropy,
it is necessary to find a non-linear solution with anisotropy. 
This deserves further investigations.

Now we discuss the behavior of $\delta \pi_F$ related to the dark
radiation, which plays an important role when we compute CMB anisotropy
in brane world models.
As shown in Sec.\ref{kk}, $\delta \pi_F$ that originates from the
perturbatively small black hole mass in the bulk is related to 
$\delta \rho_F$ at large scales as
%
\begin{eqnarray}
\delta \pi_F = \frac{(-k\eta)^2}{4\mu(\mu+1)} \delta \rho_F.
\label{stress_to_rho}
\end{eqnarray}
%

Let us compare this result with the one obtained in
\cite{Koyama} in the RS two brane model by solving the
bulk geometry using a low energy approximation \cite{ShiromizuKoyama}.
(It should be noted that we are assumed to live on the positive
tension brane.)
In \cite{Koyama}, the matter on the brane is assumed to have the
equation of state $P=w\rho$.
If we consider the four-dimensional cosmological constant as the
matter on the brane ($w=-1$), the model in
\cite{Koyama} corresponds to our model with $b=0$ except for the
existence of the second brane.
If the distance between two branes is constant, the solution for 
an anisotropic component of $E_{\mu \nu}$ becomes 
%
\begin{eqnarray}
\kappa_4^2 \delta \pi_F &=& \frac{1}{5} k^2 a^{-2} \frac{\delta
\rho_F}{\rho}
\\
&=& \frac{1}{5} k^2 a^{-2} \frac{\delta C_r a^{-4}}{\rho},
\label{lowpi}
\end{eqnarray}
%
where $a$ is the cosmic scale factor and $\kappa_4^2 \rho = 3 H^2$.
$\delta C_r$ is the integration constant associated with the
perturbation of the radion which is defined as the physical distance
between two branes.
If we rewrite the above relation using the conformal time,
the result Eq.(\ref{lowpi}) becomes
%
\begin{eqnarray}
\delta \pi_F &=& \frac{(-k\eta)^2}{15} \delta \rho_F
\end{eqnarray}
%
This relation is exactly the same as Eq.(\ref{stress_to_rho})
for $b=0 (\mu=3/2)$. 
It is somewhat a surprising result. Despite the fact that 
the low energy expansion scheme is applicable only for two 
branes model, our result shows that it can be used to 
investigate the bulk gravitational field in one brane model 
if we choose the boundary condition at the second brane properly. 
We plan a more detailed study on the effectiveness of the 
low energy expansion scheme using our exactly solvable model \cite{future}.

In this paper, we did not consider the normalization of the
perturbations. 
This can be fixed if we perform the quantization of the 
perturbations. This was partially done in \cite{KoyamaTakahashi}
for the scalar field perturbation. However, precisely speaking, 
we need to quantize two degrees of freedom independently. This issue 
is also left for a future study \cite{future}.

\section*{Acknowledgments}

The work of K.K. is supported by JSPS.

\section*{APPENDIX A: PERTURBED EINSTEIN EQUATIONS ON THE BRANE}

In this appendix, we present the perturbed effective four-dimensional
Einstein equations.
The linear scalar metric and scalar field in the longitudinal gauge is
taken as 
\begin{eqnarray}
ds^2 &=& -(1+2 \Phi(t) Y) dt^2 +e^{2\alpha}  (1+2 \Psi(t) Y) 
\delta_{ij} dx^i dx^j,
\nonumber\\
\varphi &=& \varphi + \delta \varphi Y.
\end{eqnarray}

The perturbed four-dimensional Einstein equations are given by

(t,t): 
%
\begin{eqnarray}
&&6\dot{\alpha}^2 \left( \Phi - \frac{\dot{\Psi}}{\dot{\alpha}} \right)
-2k^2 e^{-2\alpha} \Psi \nonumber\\
&&~~~= -\delta \rho_F -3\sqrt{2}b\kappa \dot{\alpha} \dot{\varphi}
\left( \Phi - \frac{\dot{\Psi}}{\dot{\alpha}} \right)
+ \sqrt{2}b\kappa k^2 e^{-2\alpha} \delta\varphi\nonumber\\
&&~~~~- 2\sqrt{2}b\kappa \frac{\Delta+4}{\Delta} \lambda_0^2
\delta e^{- 2\sqrt{2}b\kappa \varphi} \delta\varphi
\label{4Dtt_bianchi}
\end{eqnarray}
%

(i,i):
%
\begin{eqnarray}
&&\ddot{\Psi} +3\dot{\alpha}\dot{\Psi} - 2\Phi \ddot{\alpha} -
3\dot{\alpha}^2 \Phi - \dot{\alpha}\dot{\Phi} + \frac{1}{3} k^2
e^{-2\alpha} (\Psi + \Phi) \nonumber\\
&&~~~=-\sqrt{2}b\kappa \biggl((\dot{\Psi} - 2\dot{\alpha}\Phi)
\dot{\varphi} - \ddot{\varphi} \Phi -\frac{1}{2} \dot{\varphi}
\dot{\Phi} + \frac{1}{2} \delta \ddot{\varphi} \nonumber\\
&&~~~~ +\dot{\alpha} \delta\dot{\varphi}
+ \frac{1}{3} k^2 e^{-2\alpha} \delta\varphi
\biggr)
+\frac{\Delta + 11/3}{2} \kappa^2 (\dot{\varphi}^2 \Phi - \dot{\varphi}
\delta\dot{\varphi})
\nonumber\\&&~~~~
+ \sqrt{2}b\kappa \frac{\Delta+4}{\Delta}
\lambda_0^2 \delta e^{- 2\sqrt{2}b\kappa \varphi} \delta\varphi
\nonumber\\&&~~~~
-\frac{1}{6} (\delta \rho_F + 3\sqrt{2}b\kappa \delta F_\varphi)
\label{4Dii_bianchi}
\end{eqnarray}
%

(i,j):
%
\begin{eqnarray}
-k^2 e^{-2\alpha} (\Phi + \Psi + \sqrt{2}b\kappa \delta \varphi)
= \delta \pi_F,
\label{4Dij_bianchi}
\end{eqnarray}
%

(t,i):
%
\begin{eqnarray}
&&-2 ke^{-\alpha} (\dot{\Psi} - \dot{\alpha} \Phi)
= \frac{2\kappa^2}{3} ke^{-\alpha} \dot{\varphi} \delta\varphi 
\nonumber\\
&&~~~+ \sqrt{2}b\kappa ke^{-\alpha} (\delta \dot{\varphi} + \sqrt{2}b\kappa
\dot{\varphi} \delta\varphi - \dot{\varphi} \Phi)
- \delta q_F.
\label{4Dti_bianchi}
\end{eqnarray}
%

Introducing a canonical variable for scalar perturbations 
%
\begin{eqnarray}
Q=\delta\varphi -\frac{\dot{\varphi}}{\dot{\alpha}} \Psi =
\delta\varphi -\frac{3\sqrt{2}b}{\kappa} \Psi,
\label{defQ_canonical}
\end{eqnarray}
%
the perturbed scalar field equation can be rewritten as
%
\begin{eqnarray}
\ddot{Q} + (3\dot{\alpha} + \sqrt{2}b\kappa\dot{\varphi}) \dot{Q}
+ k^2 e^{-2\alpha} Q \nonumber\\
=- \sqrt{2}b\kappa^{-1} (\delta \rho_F + \delta \pi_F) - \delta
F_{\varphi}.
\label{eqQ}
\end{eqnarray}
%
Here we used the scalar field equation (\ref{eomF}) and the
four-dimensional Einstein equations (\ref{4Dtt_bianchi}),
(\ref{4Dii_bianchi}) and (\ref{4Dij_bianchi}).
It should be noted that $Q$ is related to the curvature perturbation
as
%
\begin{eqnarray}
R_c = \frac{\dot{\alpha}}{\dot{\varphi}} Q,
\end{eqnarray}
%
which affect the amplitude of the CMB anisotropy.

The above Einstein equations and the equation for $Q$ are not closed
but include the terms due to the KK modes.
As shown in the section \ref{viewbrane}, the constraint equations for
$\delta F_{\mu\nu}$ and $\delta F_\varphi$ Eq.(\ref{relationF}) are
not sufficient to determine these variables.
We must solve the bulk dynamics to completely understand cosmological
perturbations on the brane.

\section*{Appendix C: SOLUTIONS IN THE LONGITUDINAL GAUGE}
\label{solutions_longitudinal}

In this appendix, we present the solutions of five-dimensional
Einstein equations for scalar perturbation in the longitudinal gauge.
By a gauge transformation of the solutions given in
Sec.\ref{solutions} to the longitudinal gauge, we get
%
\begin{eqnarray}
&&\Psi_L = -\frac{2}{3(\Delta+2)} \frac{i\nu+\mu-1}{i\nu-1} c_1 
\nonumber\\&&~
\times \biggl[ \frac{\Delta+2}{\Delta+4} (i\nu+\mu)(i\nu-\mu) \varrho
(\eta) \psi_m - \varrho (\eta) \frac{{\rm cosh}Hy}{{\rm sinh}Hy}
\frac{\psi_m'}{H}
\nonumber\\&&~
+ \frac{2}{\Delta+4} (i\nu-1)(i\nu+\mu)
\frac{\rho}{(-k\eta)^2} \zeta_m(y) \biggr]
\nonumber\\&&~
+\frac{3\Delta+8}{3(\Delta+4)} c_2 \biggl[
\frac{i\nu+\mu-1}{i\nu-1} \varrho (\eta) \psi_m
\nonumber\\&&~
+ \frac{2}{\Delta+2}
\frac{\rho}{(-k\eta)^2} \zeta_m(y) \biggr],
\\
&&\Phi_L = \frac{2}{3(\Delta+4)} (i\nu+\mu)(i\nu+\mu-1) c_1 
\nonumber\\&&~
\times \biggl[ -\frac{1}{(i\nu+\mu)(i\nu-1)} \varrho (\eta)
\biggl( (i\nu+\mu)(i\nu-\mu) \psi_m 
\nonumber\\&&~
- \frac{\Delta+4}{\Delta+2} \frac{{\rm
cosh}Hy}{{\rm sinh}Hy} \frac{\psi_m'}{H} \biggr)
-\frac{2}{\Delta+2} \frac{\rho}{(-k\eta)^2} \frac{{\rm
cosh}Hy}{{\rm sinh}Hy} \frac{\psi_m'}{H}
\nonumber\\&&~
+(i\nu-\mu)(3i\nu+3-\mu) \frac{\rho}{(-k\eta)^2} \psi_m \biggr]
\nonumber\\&&~
-\frac{3\Delta+8}{3(\Delta+4)} c_2 \biggl[ \biggl( -\rho_{i\nu} +
\frac{4\mu-3}{2(i\nu-1)} (\rho_{i\nu}+\rho_{i\nu-2})
\nonumber\\&&~
+(i\nu-\mu)(3i\nu+3-\mu) \frac{\rho}{(-k\eta)^2} \biggr) \psi_m
\nonumber\\&&~
-\frac{2}{\Delta+2} \frac{\rho}{(-k\eta)^2} \frac{{\rm
cosh}Hy}{{\rm sinh}Hy} \frac{\psi_m'}{H} \biggr],
\\
&&N_L = -\Psi_L -\Phi_L,
\\
&&A_L = 2k^{-1} e^\alpha \psi_m' \biggl[ c_1 k \eta \mu
\frac{i\nu+\mu-1}{i\nu-1} (\rho_{i\nu} + \rho_{i\nu-2})
\nonumber\\&&~
-\frac{3\Delta+8}{\Delta+4} \biggl(c_2 - \frac{2 c_1}{3\Delta+8}
(i\nu+\mu)(i\nu+\mu-1) \biggr)
\nonumber\\&&~
\times \frac{1}{k\eta} (k\eta\rho_{i\nu-1} + (i\nu-\mu+1) \rho_{i\nu})
\biggr],
\\
&&\delta \varphi_L = 3\sqrt{2}b\kappa^{-1} (-c_2 \rho_{i\nu} + \Psi_L),
\end{eqnarray}
%
where
%
\begin{eqnarray}
\varrho(\eta) = \rho_{i\nu} + \frac{\mu}{i\nu+\mu-1} \rho_{i\nu-2},
\\
\zeta_m(y) = (i\nu-\mu) \psi_m + \frac{{\rm cosh}Hy}{{\rm sinh}Hy}
\frac{\psi_m'}{H}.
\end{eqnarray}
%
If we take 
%
\begin{eqnarray}
c_1 = \frac{c_2 (3\Delta+8)}{2(i\nu+\mu)(i\nu+\mu-1)},
\end{eqnarray}
%
this solution becomes the one already obtained in Koyama and
Takahashi \cite{KoyamaTakahashi,KoyamaTakahashi2}.

\section*{Appendix D: PERTURBATION FORMULAS FOR 5D WEYL TENSOR}
\label{perturbation_formula}

In this appendix, we give the Wey tensor in our background spacetime.
Here we take the Gaussian normal gauge condition
Eq.(\ref{gauge_cond}).
The explicit expressions are as follows;
\begin{eqnarray}
C &=& \frac{e^{2W}}{2}  \biggl[\Phi''-\Psi''\biggr]
+ \frac{ e^{2W+2\sqrt{2}b\kappa\varphi}}{2} \biggl[-\ddot{N}
+\ddot{\Psi} - \frac{k}{3} \dot{B}
\nonumber\\&&
+ \sqrt{2}b\kappa\dot{\varphi} \biggl(\dot{\Phi} +
\dot{\Psi} - \frac{k}{3} B -2\dot{N} \biggr)
+\dot{\alpha}\biggl( \dot{N} - \dot{\Phi}\biggr) 
\nonumber\\&&
- 2N ( \sqrt{2}b\kappa\ddot{\varphi} + 2b^2\kappa^2 \dot{\varphi}^2
-\ddot{\alpha} - \sqrt{2}b\kappa\dot{\varphi}\dot{\alpha})
\biggr]
\nonumber\\&&
+\frac{k^2}{6} e^{2W+2\sqrt{2}b\kappa\varphi -2\alpha}
\biggl[N+\Phi-2\Psi-\frac{2}{3}E\biggr],
\end{eqnarray}
%
%
\begin{eqnarray}
C_2 &=& -\frac{2}{3} e^{2W} E''
+\frac{k^2}{3} e^{2W+2\sqrt{2}b\kappa\varphi-2\alpha} \biggl[ -2N +
\Psi + \Phi
\nonumber\\&&
+ \frac{1}{3} E \biggr]
-\frac{1}{3} e^{2W+2\sqrt{2}b\kappa\varphi} \biggl[ \ddot{E} +
k\dot{B}
\nonumber\\&&
- (\dot{E}+kB)(2\sqrt{2}b\kappa\dot{\varphi} - 3\dot{\alpha}
)
\nonumber\\&&
+E ( \sqrt{2}b\kappa\ddot{\varphi} + 2b^2\kappa^2 \dot{\varphi}^2
-\ddot{\alpha} - \sqrt{2}b\kappa\dot{\varphi}\dot{\alpha})
\biggr],
\end{eqnarray}
%
where we expanded $\delta ^{(5)} C_{\mu y\nu y}$ in terms of the
scalar harmonics as
%
\begin{eqnarray}
&&\delta ^{(5)} C_{tyty} = C Y,
\nonumber\\
&&\delta ^{(5)} C_{iyjy} = e^{2\alpha} \left( \frac{1}{3} C Y
\delta_{ij} + C_2 Y_{ij} \right).
\end{eqnarray}
%


\end{document}